\documentclass[12pt,aps,prc,showpacs,nofootinbib]{revtex4}
\usepackage{amssymb}
\usepackage{makeidx}
\usepackage{amsmath}
\usepackage{amsfonts}
\usepackage{graphicx}

\usepackage{bm}
\usepackage{rotating}
\usepackage{epsfig}
\usepackage{amsmath}
\usepackage{amsfonts}

\begin{document}

\title{The contribution of axial-vector mesons to hyperfine\\ structure of muonic hydrogen}

\author{A.~E.~Dorokhov\footnote{E-mail:~dorokhov@theor.jinr.ru}}
\affiliation{Joint Institute of Nuclear Research, BLTP,\\
141980, Moscow region, Dubna, Russia}
\author{N.~I.~Kochelev\footnote{E-mail:~nikkochelev@mail.ru}}
\affiliation{Institute of Modern Physics of Chinese Academy of Sciences, 730000, Lanzhou, China}
\affiliation{Joint Institute of Nuclear Research, BLTP,\\
141980, Moscow region, Dubna, Russia}
\author{A.~P.~Martynenko\footnote{E-mail:~a.p.martynenko@samsu.ru}}
\affiliation{Samara University, 443086, Samara, Russia}
\author{F.~A.~Martynenko\footnote{E-mail:~f.a.martynenko@gmail.com}}
\affiliation{Samara University, 443086, Samara, Russia}
\author{A.~E.~ Radzhabov\footnote{E-mail:~aradzh@icc.ru}}
\affiliation{
Institute of Modern Physics, Chinese Academy of Sciences, Lanzhou 730000, China}
\affiliation{
Matrosov Institute for System Dynamics and Control Theory SB RAS, 664033, Irkutsk, Russia }

\begin{abstract}
The contribution from the axial-vector meson exchange to the potential of
the muon-proton interaction in muonic hydrogen induced by anomalous axial-vector meson coupling
to two photon state is calculated. It is shown that such contribution to
the hyperfine splitting in muonic hydrogen is large and important for a comparison with precise experimental data.
In the light of our result, the proton radius "puzzle" is discussed.
\end{abstract}


\maketitle

\section{Introduction}

Seven years ago, the CREMA (Charge Radius Experiments with Muonic Atoms) Collaboration \cite{crema1}
measured  very precisely the Lamb shift of muonic hydrogen. This event opened a new era
of precise investigation of the energy spectrum of simple atoms. Furthermore, in the new experiments of this Collaboration
with muonic deuterium and ions of muonic helium a charge radii of light nuclei were obtained with very high precision  \cite{crema2,crema3,crema4}. In the case of muonic hydrogen and muonic deuterium it was shown that obtained values of the charged radii
are significantly different from those which were extracted from experiments with electronic atoms and in the scattering of the electrons with
nuclei and were recommended for using by the CODATA \cite{Mohr:2012tt}. At present, several experimental groups plan 
to measure the hyperfine structure (HFS) of various muonic atoms with more high precision \cite{ma_2017,adamczak_2017,pohl_2017}.
This will make it possible to better understand the existing "puzzle" of the proton charge radius, to check the Standard Model with greater accuracy and, possibly, to reveal the source of previously unaccounted interactions between the particles forming the bound state. One way to overcome the crisis situation is a deeper theoretical analysis of the fine and hyperfine structure
of muonic atom spectrum, in the verification of previously calculated contributions and the more accurate construction of the particle interaction operator in quantum field theory, the calculation of new corrections whose value for muonic atoms can increase substantially in comparison with electronic atoms.
The expected results will allow to get also a new very important information about the forces which are
responsible for the structure of atoms. From the theory side it is urgently needed to study the possible
effects of exchanges between muon and proton which can contribute to hyperfine structure of muonic
hydrogen. One of such effects was considered in recent papers \cite{apmlet,pascalutsa,pang,kou}.
It arises from the effective pion exchange between muon and proton induced by coupling of the pion to two
photons (see Fig.~\ref{fig1}~(left)). Despite the fact that numerically such contribution was found to be rather small, 
it can be important for the interpretation of new data.

In this Letter we consider the additional contribution to hyperfine structure of muonic
hydrogen which is related to the axial-vector mesons exchanges (see Fig.~\ref{fig1}~(right)).
We would like to point out, that one can expect the important contribution of this exchange to spin dependent part of muon-proton interaction
because the exchange particle has the spin one. Furthermore, it is also well known that in the channel with quantum number $1^{++}$ axial anomaly effects can play an important role and, in particularly, these effects might be considered
as a cornerstone to solve so-called "proton spin crisis" \cite{Anselmino:1994gn}.

\section{Axial-vector meson exchange contribution to muon-proton interaction induced by axial anomaly}

\begin{figure}[th]
\centerline{
\includegraphics[scale=1.0]{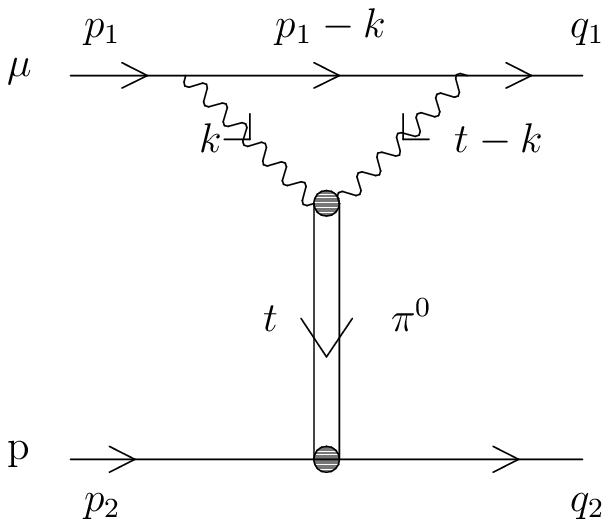}
\includegraphics[scale=1.0]{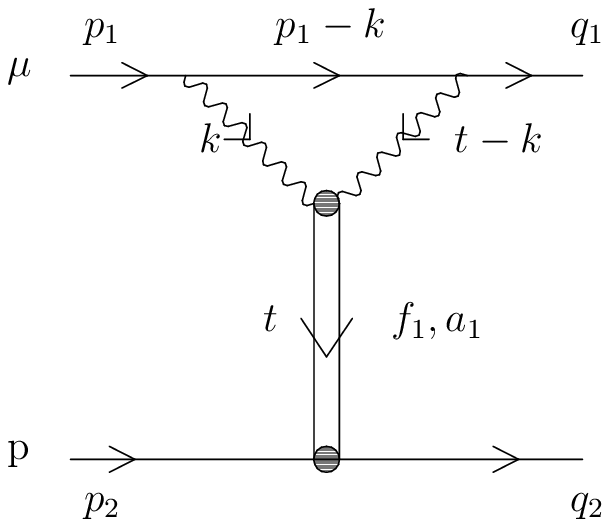}
}
\caption{Muon-proton interaction induced by mesonic exchange.}
\label{fig1}
\end{figure}

One-photon exchange interaction in quantum electrodynamics gives the leading order contribution to the 
interaction operator in muonic hydrogen. The potential of hyperfine interaction has the following form \cite{apm2004}:
\begin{equation}
\label{eq:3}
\Delta V_B^{hfs}=\frac{8\pi\alpha\mu_p}{3m_\mu m_p}({\bf S}_p{\bf S}_\mu)\delta({\bf r})
-\frac{\alpha\mu_p(1+a_\mu)}{m_\mu m_pr^3}\left[({\bf S}_p{\bf S}_\mu)-3({\bf S}_p{\bf n})
({\bf S}_p{\bf n})\right]+
\end{equation}
\begin{displaymath}
\frac{\alpha\mu_p}{m_\mu m_pr^3}\left[1+\frac{m_\mu}{m_p}-\frac{m_\mu}{2m_p\mu_p}\right]({\bf L}{\bf S}_p)
\end{displaymath}
where  $m_\mu $, ${\bf S}_\mu$  and $m_p$,  ${\bf S}_p$ are masses and spins  of muon and proton,  correspondingly, $\mu_p$  is the proton magnetic moment.
The potential (\ref{eq:3}) gives the main contribution of order $\alpha^4$
to the hyperfine structure of muonic atom.
Precision calculation of the hyperfine structure of the spectrum, which is necessary for a comparison with experimental data,
requires the consideration of various corrections to the vacuum polarization, nuclear structure and recoil,
and relativistic corrections \cite{apm2004,egs,borie,apm2017}. We calculate further the contribution to HFS which is
determined by the axial-vector $f_1(1285)$, $a_1(1260)$ and $f_1(1420)$ meson exchanges shown in Fig.~\ref{fig1} (right).

\begin{figure}[th]
\centerline{\includegraphics[scale=0.9]{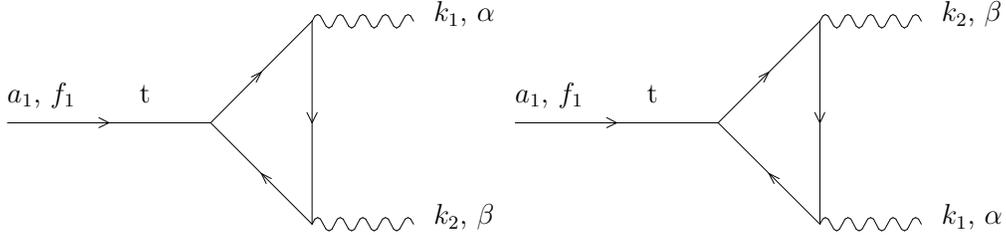}}
\caption{Coupling of axial-vector mesons to two photons.}
\label{fig2}
\end{figure}

The coupling of the axial-vector meson to two photon state is possible through anomalous
triangle diagram, shown in Fig.~\ref{fig2}.
The general structure of this vertex takes the form \cite{Rose63,ABJ69,bell}:
\begin{eqnarray}
T^{\mu\nu\alpha}&=& 4\pi i \alpha \, \varepsilon_{\rho \sigma \tau \alpha} \,  \biggl\{
\left[\nu \left(A_3 k_1^{\tau}- \tilde{A}_3k_2^{\tau}\right) + k_2^2 A_4 k_1^{\tau}
-k_1^2 \tilde{A}_4 k_2^{\tau}\right] g^{\mu \rho}g^{\sigma \nu}  \nonumber \\
&&\quad
+A_3 k_1^{\nu} k_1^{\rho} k_2^{\sigma}g^{\tau \mu}- \tilde{A}_3 k_2^{\mu} k_1^{\rho} k_2^{\sigma}g^{\tau \nu}+A_4 k_2^{\nu} k_1^{\rho} k_2^{\sigma}g^{\tau \mu}
-\tilde{A}_4 k_1^{\mu} k_1^{\rho} k_2^{\sigma}g^{\tau \nu}
\biggr\}.\label{eq:a1}
\end{eqnarray}
where $A_i\equiv A_i(t^2,k_1^2,k_2^2)$, $\tilde{A}_i\equiv A_i(t^2,k_2^2,k_1^2)$. Another form of the tensor describing the transition from initial state of two virtual photons with four-momenta $k_1$, $k_2$ to an axial-vector
meson ${\cal A}$ ($J^{PC}=1^{++}$) with the mass $M_A$ is presented in \cite{Pascalutsa:2012pr}\footnote{
The only difference between our expression \eqref{eq:aa1} and their work is related to the normalized factor $1/M_A^2$ used in \cite{Pascalutsa:2012pr}.}:
\begin{equation}
\label{eq:aa1}
T^{\mu\nu\alpha}=4\pi i\alpha\varepsilon_{\rho\sigma\tau\alpha}\Bigr[
R^{\mu\rho}(k_1,k_2)R^{\nu\sigma}(k_1,k_2)(k_1-k_2)^\tau\nu F^{(0)}_{A\gamma^\ast\gamma^\ast}(k_1^2,k_2^2)+
\end{equation}
\begin{displaymath}
+R^{\nu\rho}(k_1,k_2)\left(k_1^\mu-\frac{k_1^2}{\nu}k_2^\mu\right)k_1^\sigma k_2^\tau
F^{(1)}_{A\gamma^\ast\gamma^\ast}(k_1^2,k_2^2)+
\end{displaymath}
\begin{displaymath}
+R^{\mu\rho}(k_1,k_2)\left(k_2^\nu-\frac{k_2^2}{\nu}k_1^\nu\right)k_2^\sigma k_1^\tau
F^{(1)}_{A\gamma^\ast\gamma^\ast}(k_2^2,k_1^2)\Bigl],
\end{displaymath}
\begin{displaymath}
\nu=(k_1k_2)=\frac{1}{2}[(k_1+k_2)^2-k_1^2-k_2^2]=\frac{1}{2}[t^2-k_1^2-k_2^2],
\end{displaymath}
\begin{displaymath}
R^{\mu\nu}=R^{\nu\mu}=-g^{\mu\nu}+\frac{1}{X}\left[(k_1k_2)(k_1^\mu k_2^\nu+k_1^\nu k_2^\mu)-
k_1^2k_2^\mu k_2^\nu-k_2^2k_1^\mu k_1^\nu\right],
\end{displaymath}
\begin{displaymath}
X=(k_1k_2)^2-k_1^2k_2^2=\frac{1}{4}\left[t^4-2t^2(k_1^2+k_2^2)+(k_1^2-k_2^2)^2\right].
\end{displaymath}

The relation between $A_i$, $\tilde A_i$ and form factors in \eqref{eq:aa1} is the following:
\begin{equation}
F_{{\cal A} \gamma^\ast \gamma^\ast}^{(1)}(t^2,k_1^2, k_2^2) =\frac{\nu}{X}
\left[k_2^2(\tilde{A}_3  - A_4)+ \nu (\tilde{A}_4 - A_3)\right],
\end{equation}
\begin{displaymath}
F^{(1)}_{{\cal A} \gamma^\ast \gamma^\ast}(t^2,k_2^2, k_1^2)  =
\frac{\nu}{X} \left[k_1^2( A_3 - \tilde{A}_4 )+ \nu( A_4 -  \tilde{A}_3 )
\right]
\end{displaymath}
\begin{displaymath}
F^{(0)}_{{\cal A} \gamma^\ast \gamma^\ast}(t^2,k_1^2, k_2^2) =
\frac{1}{\nu(k_1^2-k_2^2)}\left[
(\nu A_3 + k_2^2 A_4) (k_1^2+\nu)-(k_1^2 \tilde{A}_4 + \nu  \tilde{A}_3) (k_2^2+\nu)\right].
\end{displaymath}
The form factors $F^{(0)}_{A \gamma^\ast\gamma^\ast} (k_1^2,k_2^2)$ and $ F^{(1)}_{A\gamma^\ast\gamma^\ast}(k_1^2, k_2^2)$ entering in \eqref{eq:aa1} are dependent on the squares of the 4-momenta of virtual photons.
With increasing $k_1^2$,
$k_2^2$, these functions must decrease rapidly to ensure the ultraviolet convergence of the loop integral in the interaction amplitude.

We should mention that in the opposite to the case of pion coupling to two
photons, axial-vector meson can not decay into two real photons,
according to the Landau-Yang theorem \cite{Landau:1948kw,Yang:1950rg}.
Nevertheless, the coupling of
$1^{++}$ mesons  to two photons is still possible in the case when one or both photons are virtual.
For small values of relative momenta of particles in the initial and final states and small
value of transfer momentum $t$ between muon and proton, the transition amplitude
presented in \eqref{eq:a1} takes a simple form
\begin{equation}
\label{eq:a1a}
T^{\mu\nu\alpha}=8\pi i\alpha\varepsilon_{\mu\nu\alpha\tau}k^\tau k^2
F^{(0)}_{AV\gamma^\ast\gamma^\ast}(t^2,k^2,k^2),
\end{equation}
where $k=k_1=-k_2$. To extract HFS part of the interaction in the case of the
S-states the following projection operators are used for states with spin S=0 and S=1 \cite{apm2002}:
\begin{equation}
\label{eq:5}
\hat\Pi_{S=0}[u(0)\bar v(0)]_{S=0}=\frac{1+\gamma^0}{2\sqrt{2}}\gamma_5,~~~~~
\hat\Pi_{S=1}[u(0)\bar v(0)]_{S=1}=\frac{1+\gamma^0}{2\sqrt{2}}\hat\varepsilon,
\end{equation}
where $\varepsilon^\mu$ is the polarization  vector of $^3S_1$ state.
The amplitude of the muon-proton interaction presented in Fig.~\ref{fig1} (right)
has the following structure:
\begin{equation}
\label{eq:c1}
i{\cal M}=[\bar l (q_1)\Gamma_\alpha^{(\mu)} l (p_1)]{\cal D}^{\alpha\beta}(t)
[\bar N(p_2)\Gamma_\beta^{(p)} N(q_2)],
\end{equation}
where the vertex operator in the proton line is fixed
by the Hamiltonian of nucleon-axial-vector meson interaction
\begin{equation}
H_I(a_{1NN})=g_{a_1NN}\bar N {\mathstrut\bm\tau}\gamma_\mu\gamma_5N {\bf a}_1^\mu,
\end{equation}
for $a_1$ exchange and
\begin{equation}
H_I(f_{1NN})=g_{f_1NN}\bar N\gamma_\mu\gamma_5N f_1^\mu
\end{equation}
for $f_1$ exchange.
The vertex operator in the lepton line $\Gamma_\alpha^{(\mu)}$ is fixed by the integration
over $k$ in photonic loop, and
${\cal D}^{\alpha\beta}(t)$ is the axial-vector meson propagator.
It is easy to show that
\begin{equation}
\Gamma_\alpha^{(\mu)}\sim \gamma_\alpha\gamma_5,
\end{equation}
and using the relations \cite{Anselmino:1994gn}
\begin{equation}
\bar N (P,S_p)\gamma_\tau\gamma_5N(P,S_p)=2S_p^\tau ,\ \  \bar l (q,S_\mu)\gamma_\tau\gamma_5l(q,S_\mu)=
2S_\mu^\tau ,
\end{equation}
one can see that the interaction \eqref{eq:c1} contains the spin-spin interaction,
${\cal M}\sim {\bf S}_p {\bf S}_\mu $ which contributes to hyperfine splitting.
Performing the projection of the amplitude \eqref{eq:c1} to the two particle states with the help of
\eqref{eq:5}, we obtain that the numerator of the one-meson exchange amplitude (see Fig.~\ref{fig1} (right))
contains a trace of the product of the Dirac gamma-matrices and numerous convolutions by the Lorentz indices:
\begin{equation}
\label{eq:a3}
{\cal N_A}=8\pi i\alpha\varepsilon_{\rho\sigma\tau\alpha}g^{\mu\rho}g^{\nu\sigma}k^\tau k^2
g^{\alpha\beta}F^{(0)}_{AV\gamma^\ast\gamma^\ast}(t^2,k^2,k^2)g_{AVNN} \times
\end{equation}
\begin{displaymath}
Tr\left[
(\hat q_1+m_\mu)\gamma^\nu(\hat p_1-\hat k+m_\mu)\gamma^\mu(\hat p_1+m_\mu)\frac{1+\gamma_0}{2\sqrt{2}}
\hat\varepsilon(\hat p_2-m_p)\gamma_\beta \gamma_5(\hat q_2-m_p)\hat\varepsilon^\ast\frac{1+\gamma_0}{2\sqrt{2}}
\right],
\end{displaymath}
where $p_{1,2}$ ($q_{1,2}$) are initial (final) momenta of the muon and proton.
For the case of spin zero state the substitution $\hat\varepsilon\to \gamma_5$ should be done in \eqref{eq:a3}.
Introducing the total and relative momenta of particles in the initial and final states
$p=(0,{\bf p})$ and $q=(0,{\bf q})$, and taking into account their smallness in the bound state
($|{\bf p}|\sim\mu\alpha$, $|{\bf q}|\sim\mu\alpha$) ($\mu$ is the reduced mass), we can obtain the leading order contribution to
${\cal N_A}$ which does not have terms proportional to the powers of transfer momentum
$t=p-q$. Our result for hyperfine part of the potential is the following:
\begin{equation}
\label{eq:a4}
\Delta V^{hfs}_{AV}({\bf p}-{\bf q})=-\frac{32\alpha^2g_{AVpp}}{3\pi^2({\bf t}^2+M^2_A)}
\int id^4k \frac{(2k^2+k_0^2)}{k^2(k^2-2m_\mu k_0)}F^{(0)}_{AV\gamma^\ast\gamma^\ast}(0,k^2,k^2).
\end{equation}

After an analytical integration in \eqref{eq:a4} over angular variables more simple formula for
the potential can be obtained:
\begin{equation}
\label{eq:a5}
\Delta V^{hfs}_{AV}({\bf p}-{\bf q})=-\frac{32\alpha^2g_{AVpp}}{3\pi^2({\bf t}^2+M^2_A)}
\int\limits_0^\infty d k^2 L_\mu(k^2) F^{(0)}_{AV\gamma^\ast\gamma^\ast}(0,k^2,k^2),
\end{equation}
\begin{displaymath}
L_\mu(k^2) =\frac{\pi^2}{8m_\mu^4}\left[ k^2 (k^2 - 6 m_\mu^2) - (k^2 - 8 m_\mu^2)
\sqrt{k^2(k^2 + 4 m_\mu^2)}\right],
\end{displaymath}
where the kernel $L_\mu(k^2)$ behaves as $\sim2\pi^2\sqrt{k^2}/m_\mu$ for small $k^2$
while the asymptotic value for large $k^2$ is $9\pi^2/4$. Therefore
$L_\mu(k^2)$ effectively suppresses the region of small $k^2$.

\section{Model estimations}

One of the main ingredients in \eqref{eq:a4} is the form factor of transition of $1^{++}$
meson to two photons $F_{AV\gamma^\ast\gamma^\ast}(t^2,k^2,k^2)$.
Unfortunately, at present we have only few experimental data on it \cite{L3C,L3Ca,aihara}.
In the paper \cite{L3C} of the L3 Collaboration the reaction
$e^+e^- \to e^+e^-\gamma^\ast\gamma^\ast \to e^+e^-f_1(1285)\to e^+e^- \eta \pi^+\pi^-$
was studied and $f_1(1285)$ transition form factor was measured for the case when one of the photons
is real and another one is virtual.
In \cite{L3Ca} the production of $f_1(1420)$ was investigated by the same Collaboration in the reaction
$\gamma^\ast\gamma^\ast\rightarrow K_S^0K^{\pm}\pi^\mp$.
Using these data, we can parameterize the transition form factor for the case of two photons with equal
virtualities as
\begin{equation}
\label{eq:n1}
F^{(0)}_{AV\gamma^\ast\gamma^\ast}(M_{A}^2,k^2,k^2)=F^{(0)}_{AV\gamma^\ast\gamma^\ast}(M_{A}^2,0,0)F^2_{AV}(k^2),
\end{equation}
where
\begin{equation}
\label{eq:n16}
F_{AV}(k^2)=\frac{\Lambda_A^4}{(\Lambda_A^2-k^2)^2}.
\end{equation}

It should be mentioned that in comparison with the case of light $\pi^0$ exchange, the effects of off-shellness
for the exchange by massive $f_1$ mesons might be important.
The effect of off-shellness was investigated in \cite{dorokhov1,dorokhov3}, and in \cite{ls} a simple parametrization
was proposed. The simplest way to take it into account is to introduce an exponential suppressive factor \cite{ls}:
\begin{equation}
\label{eq:n6}
\frac
{F^{(0)}_{AV\gamma^\ast\gamma^\ast}(t^2,0,0)}
{F^{(0)}_{AV\gamma^\ast\gamma^\ast}(M_{A}^2,0,0)}
\approx e^{(t^2-M_{A}^2)/M_{A}^2},
\end{equation}
which gives the factor $\sim e^{-1}$ for $t^2\approx 0$.
The values of the form factors in \eqref{eq:n1} for the case of $f_1(1285)$ and $f_1(1420)$ can be fixed from the L3 
data using the relations given by the nonrelativistic quark model \cite{Pascalutsa:2012pr}:
\begin{equation}
\label{eq:n7}
F^{(0)}_{AV\gamma^\ast\gamma^\ast}(M_{A}^2,0,0)=-F^{(1)}_{AV\gamma^\ast\gamma^\ast}(M_{A}^2,0,0)
\end{equation}
and
\begin{equation}
\label{eq:n8}
\tilde\Gamma_{\gamma^\ast\gamma^\ast}(AV)=\frac{\pi\alpha^2M^5_{A}}{12}[F^{(1)}_{AV\gamma^\ast\gamma^\ast}(M_{A}^2,0,0)]^2,
\end{equation}
where $\tilde\Gamma_{\gamma^\ast\gamma^\ast}(AV)$ is the decay width of axial-vector meson. We would like to mention that 
according to the nonrelativistic quark model the sign of $F^{(0)}_{AV\gamma^\ast\gamma^\ast}(M_{A}^2,0,0)$ 
should be positive \cite{cahn}. Finally, we obtain from the L3 data:
\begin{eqnarray}
\label{eq:n9}
&&F^{(0)}_{f_1(1285)\gamma^\ast\gamma^\ast}\left(M_{f_1(1285)}^2,0,0\right)=(0.266 \pm 0.043)~\mathrm{GeV}^{-2}, \nonumber\\
&&F^{(0)}_{f_1(1420)\gamma^\ast\gamma^\ast}\left(M_{f_1(1420)}^2,0,0\right)=(0.193 \pm0.041)~\mathrm{GeV}^{-2}.
\end{eqnarray}
The value of form factor $ F^{(0)}_{f_1(1285)\gamma^\ast\gamma^\ast}(M_{f_1(1285)}^2,0,0)$
can be estimated also within nonrelativistic quark model \cite{cahn} using the relation
\begin{equation}
\label{eq:n2}
F^{(0)}_{f_1\gamma^\ast\gamma^\ast}(M_{f_1}^2,0,0)=24<e_q^2>R'(0)
\frac{\sqrt{2}}{\sqrt{\pi}M_{A}^{9/2}},
\end{equation}
where $R'(0)$ is the derivative of the radial wave function at the origin, $<e_q^2>$  is effective quark 
charge squared in the bound state. For the isospin $I=1$ state
$(u\bar u-d\bar d)/\sqrt{2}$ ($a_1$ meson) we have $<e_q^2>=\sqrt{2}/6$,  and for the isosinglet state
$(u\bar u+d\bar d)/\sqrt{2}$ ($f_1$ meson) $<e_q^2>$=$5\sqrt{2}/18$.
The value of $R'(0)$ can be estimated from the decay width $f_2(1270)\to\gamma+\gamma$ \cite{PDG} $3.034$~keV  
by means of the expression
\begin{equation}
\label{eq:n2}
\Gamma(f_2(1270)\to\gamma^\ast\gamma^\ast)=\frac{576}{5}\alpha^2<e_q^2>\frac{|R'(0)|^2}{M_A^4},
\end{equation}
assuming that the radial wave functions for $f_1(1285)$ and $f_2(1270)$ at the origin are the same.
The equation \eqref{eq:n2}  leads to $R'(0)\approx 0.099~GeV^{5/2}$ and
\begin{equation}
\label{eq:n10}
F^{(0)}_{f_1(1285)\gamma^\ast\gamma^\ast}(M_{f_1(1285)}^2,0,0)\approx 0.240~GeV^{-2},
\end{equation}
which is very close to the L3 value \eqref{eq:n9}. Therefore, one can believe that
nonrelativistic quark model describes the dynamics of axial-vector mesons rather well. However,   
below we will use the L3 value for $F^{(0)}_{f_1(1285)\gamma^\ast\gamma^\ast}$ form factor to decrease
the dependence of our predictions from the model.
Unfortunately, there is no data for $a_1(1260)$ meson production in $\gamma^\ast\gamma^\ast$ collisions.
We estimate $F^{(0)}_{a_1\gamma^\ast\gamma^\ast}(M_{a_1}^2,0,0)$ using the diagram presented in Fig.~\ref{fig2}
and introducing the value of quark-meson couplings $g_{a_1qq}$ and $g_{f_1qq}$.
The chiral symmetry gives the relation $g_{a_1qq}=g_{f_1qq}$ (see, for example \cite{Osipov:2017ray}). Finally,
the ratio of $a_1(1260)$ and $f_1(1285)$ form factors in this case should be equal to the ratio of the effective 
quark charges squared for $f_1(1285)$ and $a_1(1260)$:
\begin{equation}
\label{eq:ratio}
\frac{F^{(0)}_{a_1(1260)\gamma^\ast\gamma^\ast}(M_{a_1(1260)}^2,0,0)}{F^{(0)}_{f_1(1285)\gamma^\ast\gamma^\ast}(M_{f_1(1285}^2,0,0)}
\approx \frac{3}{5}.
\end{equation}
Then we obtain from \eqref{eq:n9} and \eqref{eq:ratio}:
\begin{equation}
\label{a1ff}
F^{(0)}_{a_1(1260)\gamma^\ast\gamma^\ast}(M_{a_1(1260)}^2,0,0)\approx 0.160~ GeV^{-2}.
\end{equation}

Our potential \eqref{eq:a5} of hyperfine interaction can be rewritten in the form:
\begin{equation}
\label{eq:a15}
\Delta V^{hfs}_{AV}({\bf p}-{\bf q})=-\frac{32\alpha^2g_{AVpp}F^{(0)}_{AV\gamma^\ast\gamma^\ast}(0,0,0)}
{3\pi^2({\bf t}^2+M^2_A)}I\left(\frac{m_\mu}{\Lambda_A}\right),
\end{equation}
where $I\left({m_\mu/\Lambda_A}\right)$ is a convolution of the kernel $L_\mu(k^2)$ and
form-factor $F^2_{A}(k^2)$ which are dependent on the muon mass $m_\mu$ and hadron scale
$\Lambda_A$ correspondingly ($a_\mu=2m_\mu/\Lambda_A$):
\begin{equation}
\label{eq:a6b}
I\left(\frac{m_\mu}{\Lambda_A}\right)=-\int\limits_0^\infty d k^2 L_\mu(k^2) F^2_{A}(k^2)=
-\frac{\pi^2\Lambda_A^2}{4(1-a_\mu^2)^{5/2}}\left[3\sqrt{1-a_\mu^2}-a_\mu^2(5-2a_\mu^2)\ln\frac{1+\sqrt{1-a_\mu^2}}{a_\mu}\right].
\end{equation}

Making the Fourier transform of \eqref{eq:a15} and averaging the obtained expression with the wave functions of
the $1S$ and $2S$ states, we obtain the following contribution to hyperfine splitting coming from the axial-vector exchange:
\begin{equation}
\label{eq:a6}
\Delta E^{hfs}_{AV}(1S)=\frac{32\alpha^5\mu^3g_{AVpp}F^{(0)}_{AV\gamma^\ast\gamma^\ast}(0,0,0) }
{3M_A^2\pi^3
\Bigl(1+\frac{2W}{M_A}\Bigr)^2}I\left(\frac{m_\mu}{\Lambda_A}\right),
\end{equation}
\begin{equation}
\label{eq:a6a}
\Delta E^{hfs}_{AV}(2S)=\frac{2\alpha^5\mu^3g_{AVpp}F^{(0)}_{AV\gamma^\ast\gamma^\ast}(0,0,0)
\left(2+\frac{W^2}{M_A^2}\right)}
{3M_A^2\pi^3  \Bigl(1+\frac{W}{M_A}\Bigr)^4}I\left(\frac{m_\mu}{\Lambda_A}\right),
\end{equation}
where $W=\mu \alpha$ and $\mu$ is the reduced mass.

For numerical estimate we fix the slope of form factors according to the L3 data to $\Lambda_{f_1(1285)}=1.040\pm 0.078$ GeV \cite{L3C}
and $\Lambda_{f_1(1420)}=0.926\pm 0.078$ GeV \cite{L3Ca}, and assume that $\Lambda_{a_1(1260)}\approx  \Lambda_{f_1(1285)}$.
Unfortunately, there is no direct experimental data on the value of axial-vector meson couplings to the quarks and proton.
Therefore, we estimate them using nonrelativistic quark model with chiral symmetry. One of the examples of such model is
the NJL model  \cite{Osipov:2017ray}.
Within this model the form factor of $f_1(1285)$ meson can be obtained by the calculation of triangle diagram presented in Fig.~\ref{fig2}:
\begin{equation}
\label{eq:anomaly}
F^{(0)}_{f_1(1285)\gamma^\ast\gamma^\ast}(M_{f_1(1285)}^2,0,0)=\frac{5g_{f_1(1285)qq}}{72\pi^2m^2},
\end{equation}
where m is the dynamical quark mass related to the spontaneous chiral symmetry breaking.

Another couplings are related to each others by using chiral symmetry and $SU(6)$-model for wave function 
of the proton as follows:
\begin{equation}
\label{eq:coupl1}
 g_{a_1(1260)qq}=g_{f_1(1285)qq},  ~~~g_{f_1(1285)pp}=g_{f_1(1285)qq}, ~~~g_{a_1(1260)pp}=\frac{5}{3}g_{f_1(1285)qq}.
\end{equation}
In the most versions of quark models which are used in hadron spectroscopy, the value of quark mass is in the interval
$m\sim 0.25\div 0.35$ GeV. At the central value $m=0.300$ GeV we get the following couplings:
\begin{equation}
\label{eq:coupl2}
g_{a_1(1260)qq}=g_{f_1(1285)qq}=g_{f_1(1285)pp}=3.40\pm 1.19, ~~~g_{a_1(1260)pp}=5.67\pm 1.98.
\end{equation}
The error in determining the interaction constants, which is at least 35 percent, is written out directly in \eqref{eq:coupl2}.
In the case of $f_1(1420)$ meson one should take into account the singlet-octet mixing effects \cite{L3Ca},\cite{Close:1997nm}.
The estimation given in \cite{Close:1997nm} shows that the wave function of this meson in flavour space is equal
\begin{equation}
f_1(1420)\approx |s\bar s>+\delta|n\bar n>, 
\label{f11420}
\end{equation}
where $n\bar n=\frac{1}{\sqrt{2}}(u\bar u+d\bar d)$ and $\delta\approx 0.4\div 0.5 $.
Therefore,  for the proton wave function in the OZI limit one can neglect the interaction of strange component 
of $f_1(1420)$ with proton and obtain the following estimation:
\begin{equation}   
g_{f_1(1420)pp}\approx 1.36\div 1.70.
\label{f11420coupl}
\end{equation}
The central value $g_{f_1(1420)pp}=1.51$ is taken for numerical estimate.

Our results for the contribution of the axial-vector mesons to HFS  are presented in Table~\ref{tb1}.
For the case of both $1S$ and  $2S$ states, the summary contribution of axial-vector meson exchanges 
is more than an order of magnitude greater than the contribution
of pseudoscalar mesons and very important to obtain the total value of the HFS with high precision.
We can use the obtained expressions \eqref{eq:a6}-\eqref{eq:a6a} to estimate the similar contribution to the 
hyperfine structure of electronic hydrogen. In the case of the 1S state, the total contribution 
of the axial vector mesons $f_1$, $a_1$ is about 0.8 kHz, which is comparable with the error 
in calculating the contribution to the proton polarizability \cite{fm2002}.

\begin{table}[h]
\caption{\label{tb1} Axial-vector meson contributions to hyperfine structure of muonic hydrogen.}
\bigskip
\begin{tabular}{|c|c|c|c|c|c|c|}   \hline
AV meson & $I^G(J^{PC})$  & $\Lambda_A$ & $g_{AVpp}$& $F^{(0)}_{AV\gamma^\ast\gamma^\ast}(0,0)$  & $\Delta E^{hfs}(1S)$  & $\Delta E^{hfs}(2S)$  \\
   &        &   in GeV  &    & in  GeV$^{-2}$    &   in meV   &  in meV   \\  \hline
$f_1(1285)$  & $0^+(1^{++})$    & 1.040&3.40 & 0.266  &$-0.0090\pm 0.0033$   &   $-0.0011\pm 0.0004$     \\     \hline
$a_1(1260)$  & $1^-(1^{++})$ &1.040   &5.67  &  0.160  & $-0.0094\pm 0.0038$     &$-0.0012\pm 0.0005$       \\     \hline
$f_1(1420)$  & $0^+(1^{++})$   &0.926  &1.51  &  0.193  & $-0.0019\pm 0.0011$    & $-0.0002\pm 0.0001$       \\     \hline
\end{tabular}
\end{table}

\section{Conclusion}

A new important contribution to the muon-nucleon interaction is discovered. 
It is determined by the effective axial-vector meson exchange induced
by anomalous axial-vector meson vertex  with two photon state.
The contribution of this exchange to hyperfine structure of muonic hydrogen is calculated in the framework
of quasipotential method in quantum electrodynamics and with the use of the
technique of projection operators on states of two particles with a definite spin.
It is shown that this contribution is large and should be taking into account for the interpretation of
new data on HFS in muonic hydrogen.
As has been mentioned above the CREMA Collaboration measured two transition frequencies in
muonic hydrogen  for the 2S triplet state $(2P_{3/2}^{F=2}-2S_{1/2}^{F=1})$
and for the 2S singlet state $(2P_{3/2}^{F=1}-2S_{1/2}^{F=0})$ \cite{crema2}. From these measurements
it is possible to extract the value of hyperfine splitting of the $2S$ level. The obtained value $\Delta E^{hfs}_{exp}(2S)=22.8089(51)$
meV allows to get the value of the Zemach radius with accuracy
$3.4~\%$ $r_Z=1.082(31)^{exp}(20)^{th}$ with the help of following relation: $\Delta E^{hfs}_{th}=22.9843(15)-0.1621(10) R_Z$.
This is in the agreement with another numerical values  $r_Z=1.086(12)$ fm \cite{p1},
$r_Z = 1.045(4)$ fm \cite{p2}, $r_Z = 1.047(16)$ fm \cite{p3}, $r_Z = 1.037(16)$ fm \cite{p4}
obtained from electron-proton scattering and from hydrogen and muonium spectroscopy.
At present the theory estimates of hadronic corrections to the 1S hyperfine splitting
in muonic hydrogen are known with a precision near 400 ppm \cite{crema3} (see more detailed analysis 
in a recent paper \cite{tomalak}).
We should emphasize that the changing the theoretical value of the HFS on
0.001 meV  leads to the changing of the Zemach radius on
0.006 fm. Therefore, our contribution coming from axial-vector meson exchange leads to new value 
of the radius $R_Z=1.067(37)$ fm, which is greater in the comparison with most listed results 
but still agree with them within errorbars.

The CREMA Collaboration have performed successively several experiments with muonic hydrogen.
In the first experiment of 2010 \cite{crema1}, the frequency of a single $2P_{3/2}^{F=2}-2S_{1/2}^{F=1}$
transition was measured.
To extract a new value of the proton charge radius in this case, there was used the theoretical
expression for the hyperfine splitting of the 2S-level in the form:
\begin{equation}
\label{eq:ls}
\Delta E_{th}(2P_{3/2}^{F=2}\div 2S_{1/2}^{F=1})=209.9779(49)-5.2262 r_p^2+0.0347 r_p^3.
\end{equation}
Since in this paper we are just calculating the hyperfine structure of the spectrum,
our result could be related to the correction of the proton charge radius.
But already in the experiment of 2013 \cite{crema2} two transition frequencies were measured,
which made it possible to find the experimental value of the hyperfine splitting of the 2S-level.
The theoretical result was used for the hyperfine structure of the P-levels.
Since the hyperfine splitting of the 2S-level can now be considered fixed from the experiment
and used further, the theoretical contribution to the HFS obtained in this paper does not lead
to a change in the proton charge radius, which remains equal to $r_p=0.84087(39)$  fm \cite{crema2} and differs from
the value recommended by CODATA-2014, $r_p=0.8751(61) $ fm \cite{Mohr:2012tt}, based on H spectroscopy
and electron-proton scattering.

There are a number of uncertainties related to the main used parameters, among which
the key role are played by the value of form factor $F^{(0)}_{AV\gamma^\ast\gamma^\ast}(0,0)$ (or $R'(0)$) and 
coupling constants $g_{AVpp}$.
The errors in determining the parameters $\tilde\Gamma_{\gamma \gamma} $ and
$ \Lambda_A $, through which $R'(0)$ is expressed, are $25~\% $ and $ 8~\% $, respectively.
Another error of about $35~\% $ for the mesons $f_1(1285) $, $ a_1(1260 $) and $ f_1(1420) $
is related to the magnitude of the interaction constants of the axial vector mesons with the nucleon.
Therefore, from the experimental data and model approximations for constructing the transition
form factor and the interaction potential of particles, we estimate the error in calculating
the contribution of the axial vector mesons $ f_1 (1285) $ to $ 35~\% $, $ a_1(1260)$ 
to $ 40~\% $, and the meson $ f_1(1420) $ in $ 60~\% $.
All theoretical errors are directly indicated in Table~\ref{tb1}.

It is necessary to mention that our estimations are mainly based on the data of the L3 Collaboration
on the transition form factors
of the axial vector mesons in photon-photon interaction. These data are restricted by rather small
kinematical region.
A new, more detailed measurements of these form factors are urgently needed.
Such type experiment is possible, for example by the BESIII and BELLEII Collaborations.

We also believe that it is important to investigate this new contribution to the hyperfine
structure of the muonic deuterium and muonic helium.
In this case it might be even possible to separate contributions coming from $a_1$ and $f_1$ mesons due to
different isospin structure of these nuclei. The research in this direction is in progress.

\begin{acknowledgments}
The authors are grateful to R.~Pohl for careful reading of our manuscript and
useful remarks. The work is supported by Russian Science Foundation (grant No. RSF 15-12-10009) (A.E.D.),
the Chinese Academy of Sciences visiting professorship for senior international scientists
(grants No. 2013T2J0011) (N.I.K.) and President's international fellowship initiative
(Grant No. 2017VMA0045) (A.E.R.), Russian Foundation for Basic Research (grant No. 16-02-00554) (A.P.M., F.A.M.).
\end{acknowledgments}

\end{document}